%% file: main.tex
\documentclass[sigconf]{acmart}
\AtBeginDocument{%
  }

\copyrightyear{2024}
\acmYear{2024}
\setcopyright{rightsretained}
\acmConference[CIKM '24]{Proceedings of the 33rd ACM International Conference on Information and Knowledge Management}{October 21--25, 2024}{Boise, ID, USA}
\acmBooktitle{Proceedings of the 33rd ACM International Conference on Information and Knowledge Management (CIKM '24), October 21--25, 2024, Boise, ID, USA}
\acmDOI{10.1145/3627673.3679617}
\acmISBN{979-8-4007-0436-9/24/10}

\makeatletter
\gdef\@copyrightpermission{
  \begin{minipage}{0.3\columnwidth}
   \href{https://creativecommons.org/licenses/by/4.0/}{\includegraphics[width=0.90\textwidth]{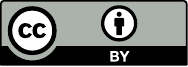}}
  \end{minipage}\hfill
  \begin{minipage}{0.7\columnwidth}
   \href{https://creativecommons.org/licenses/by/4.0/}{This work is licensed under a Creative Commons Attribution International 4.0 License.}
  \end{minipage}
  \vspace{5pt}
}
\makeatother

\usepackage{xcolor}
\usepackage[ruled,linesnumbered]{algorithm2e}
\usepackage{graphicx}
\usepackage{subfig}
\usepackage{threeparttable}
\usepackage{booktabs}
\usepackage{multirow}
\usepackage{caption}
\usepackage{url}
\usepackage{diagbox}
\usepackage{svg}
\usepackage{enumitem}
\begin{document}
\title{Watermarking Recommender Systems}

\author{Sixiao Zhang}
\affiliation{%
  \institution{Nanyang Technological University}
  \city{Singapore}
  \country{Singapore}
}
\email{sixiao001@e.ntu.edu.sg}

\author{Cheng Long}
\authornote{Co-corresponding authors.}
\affiliation{%
  \institution{Nanyang Technological University}
  \city{Singapore}
  \country{Singapore}
}
\email{c.long@ntu.edu.sg}

\author{Wei Yuan}
\affiliation{%
  \institution{The University of Queensland}
  \city{Brisbane}
  \country{Australia}
  }
\email{w.yuan@uq.edu.au}

\author{Hongxu Chen}
\affiliation{%
  \institution{The University of Queensland}
  \city{Brisbane}
  \country{Australia}
  }
\email{hongxu.chen@uq.edu.au}

\author{Hongzhi Yin}
\authornotemark[1]
\affiliation{%
  \institution{The University of Queensland}
  \city{Brisbane}
  \country{Australia}
}
\email{h.yin1@uq.edu.au}

\renewcommand{\shortauthors}{Sixiao Zhang, Cheng Long, Wei Yuan, Hongxu Chen, \& Hongzhi Yin}
\begin{abstract}
Recommender systems embody significant commercial value and represent crucial intellectual property. However, the integrity of these systems is constantly challenged by malicious actors seeking to steal their underlying models. Safeguarding against such threats is paramount to upholding the rights and interests of the model owner. While model watermarking has emerged as a potent defense mechanism in various domains, its direct application to recommender systems remains unexplored and non-trivial. In this paper, we address this gap by introducing Autoregressive Out-of-distribution Watermarking (AOW), a novel technique tailored specifically for recommender systems. Our approach entails selecting an initial item and querying it through the oracle model, followed by the selection of subsequent items with small prediction scores. This iterative process generates a watermark sequence autoregressively, which is then ingrained into the model's memory through training. To assess the efficacy of the watermark, the model is tasked with predicting the subsequent item given a truncated watermark sequence. Through extensive experimentation and analysis, we demonstrate the superior performance and robust properties of AOW. Notably, our watermarking technique exhibits high-confidence extraction capabilities and maintains effectiveness even in the face of distillation and fine-tuning processes.
\end{abstract}

%
%

\ccsdesc[500]{Information systems~Recommender systems}
\ccsdesc[300]{Information systems~Data mining}
\ccsdesc[300]{Computing methodologies~Neural networks}

\keywords{recommender systems, model watermarking, distillation}

\maketitle
\sloppy
\input{intro}
\input{related-work}

\input{methodology}

\input{experiments}

\section{Conclusion}\label{sec:con}
In this paper, we propose Autoregressive Out-of-distribution Watermarking (AOW) for watermarking recommender systems. We generate an entire sequence as the watermark and train the model to memorize it. The watermark is evaluated by doing next-item prediction given the truncated watermark sequences. The watermark sequence is generated autoregressively. We first select an initial item, then query the model to obtain scores for all items, and choose a random item from the bottom-M items with least scores. We conduct extensive experiments to demonstrate the effectiveness of AOW in protecting model ownership as well as preserving model utility, and provide comprehensive analysis on the choice of hyperparameters. Applications for AOW on federated recommendation as a fingerprinting technique would be explored in the future.

\begin{acks}
This research is supported by the Ministry of Education, Singapore, under its Academic Research Fund (Tier 2 Award MOE-T2EP20221-0013, Tier 2 Award MOE-T2EP20220-0011, and Tier 1 Award (RG77/21)). Any opinions, findings and conclusions or recommendations expressed in this material are those of the author(s) and do not reflect the views of the Ministry of Education, Singapore. This work is partially supported by Australian Research Council under the streams of Future Fellowship (Grant No. FT210100624), Linkage Project (Grant No. LP230200892), Discovery Project (Grants No. DP240101108).
\end{acks}

\balance
\bibliographystyle{ACM-Reference-Format}
\bibliography{ref.bib}
\end{document}

%% file: intro.tex
\section{Introduction}
Machine learning models have become a necessary component in the modern information society. One of the most representative machine learning products is the recommender system \cite{rendle2012bpr,zhang2019deep,he2020lightgcn,wu2022graph,long2024physical}. It has been widely deployed in various domains including health care \cite{yue2021overview}, finance \cite{sharaf2022survey}, e-commerce \cite{schafer2001commerce}, social platforms \cite{fan2019graph}, etc. It plays a crucial role in improving user experience and gaining profits for the service provider. As highly commercialized products, many of them involve certain intellectual property and copyright as well as business secrets. Some recommender systems are developed by the service provider themselves, where cutting-edge techniques might have been adopted to outperform competitors. In this case, model theft and model leakage are intolerable. Some other recommender systems are developed by a third party at the request of the service provider, where the illegal use and redistribution of the model are typically prohibited. Thus, it is necessary to take protective measures to ensure the recommender systems are not being used for these malicious aims.

To the best of our knowledge, there are a limited number of works addressing the above issue on recommender systems. Zhang et al. \cite{zhang2024defense} proposed an optimization-based method against model extraction attacks on recommender systems. They train the target model with an additional loss that can cause malfunctions to the attacker's model. However, this method inevitably degrades the performance of the target model. As a commercial product, any degradation in performance for the recommender system may lead to a huge profit loss. Thus, it is necessary to design a protection method without sacrificing the model utility.

One promising defense strategy in this scenario is the model watermarking. It has been extensively studied in computer vision \cite{uchida2017embedding,chen2019deepmarks,wang2020watermarking,zhang2018protecting,adi2018turning,guo2018watermarking,zhu2020secure,le2020adversarial,liu2021secure,atli2020waffle,yang2022watermarking}. The goal is to verify the ownership of a model by encrypting a specific pattern. For example, white-box watermarking methods assume that the parameters of a suspicious model can be accessed, thus they encode a certain message into the target model's parameters and decode it with a fixed neural network \cite{uchida2017embedding}; black-box watermarking methods assume that the model owner can only query the suspicious model without access to its parameters, thus they encode a backdoor into the model by forcing the target model to memorize a special input-output mapping \cite{adi2018turning}. If a model is suspected to be an illegal copy of the target model, the model owner can check the existence of the watermark to claim the ownership. Model watermarking does minimal harm to the model utility, because it does not interfere with the regular training process. Due to the natural gap between the data structure in computer vision and recommender systems, as well as the gap between the classification task and the ranking task, the watermarking methods in computer vision cannot be applied to recommender systems. Therefore, we aim to fill this gap and propose a model watermarking technique for recommender systems. In this paper, we focus on black-box watermarking, because compared with white-box watermarking, it is more challenging and practical. 

There are several unique challenges when designing a black-box watermarking method for recommender systems. First, unlike classification tasks where we can assign a predefined class to the watermark queries, it is non-trivial to define how the output ranking should be like for a watermark query. Since the output is a top-k ranking list, it is possible to encode various forms of watermarks. For example, we can predefine a fixed sub-sequence and insert it in the output, or we can choose a subset of items and force them to always rank higher than another subset, or we can simply promote or demote an item, etc. The best watermark design should take the task-specific constraints and objectives into consideration. Second, the watermark should not exist in an oracle model that is trained only on the regular data. Previous works use out-of-distribution (OOD) data, such as Gaussian noise images \cite{adi2018turning} and white pixels \cite{guo2018watermarking}, to define watermarks. OOD data is not presented in the regular training set, so it is unable to detect watermarks from it. Unfortunately, due to the different data structure (images v.s. sequences), these OOD forms cannot be used for recommender systems. Third, the watermark should be resistant to removal attacks such as distillation and fine-tuning. The attackers aware of the existence of the watermark would probably conduct such attacks to remove it. A valid watermark should remain detectable even after these attacks.

To this end, we propose a model watermarking method for recommender systems, namely Autoregressive Out-of-distribution Watermarking (AOW). We focus on inductive recommender systems since it is more powerful and commonly deployed in real-world scenarios, while they are also greatly threatened by distillation attacks \cite{yue2021black}. We mainly consider the sequential recommendation scenario. We generate a watermark sequence and train the model to memorize this sequence. The watermark is evaluated by asking the model to predict the next item given the former items. This watermark design is simple and effective to avoid any trivial modifications that may ruin the watermark, such as shuffling and swapping, where the attacker may take such methods to avoid any watermark that potentially exists. We generate the watermark sequence autoregressively. Specifically, we first train an oracle model with the regular training set. Then we select an item as the initial item of the watermark sequence, and query the oracle model with this single item to get the predicted scores for all items. We randomly choose one item from those with lowest scores as the next item of the watermark sequence. We repeat this process until the sequence reaches a predefined length. This autoregressive method can guarantee that the oracle model has zero accuracy on the watermark, as long as the set of items with top-k scores does not overlap with the set of items with lowest scores that we consider for generating the watermark sequence. Besides, AOW is resistant to distillation and fine-tuning as shown in the experiment section. We can still extract the watermark with a  high confidence after both attacks. Our code is publicly available\footnote{\url{https://github.com/RinneSz/AOW}}. We summarize our contributions as follows:
\begin{itemize}
    \item We propose Autoregressive Out-of-distribution Watermarking (AOW). To the best of our knowledge, it is the first work exploring model watermarking for recommender systems.
    \item AOW has a high success rate as well as the ability to maintain the utility of the model. It is also resistant to distillation and fine-tuning. 
    \item We conduct extensive experiments to show the superior performance of AOW. In the default setting, we can extract the watermark with 1.0 Recall@1 from the target model across four datasets, while achieving >0.75 Recall@10 against distillation and fine-tuning. Comprehensive analysis on the choices of hyperparameters are presented.
\end{itemize}


%% file: related-work.tex
\section{related work}
\subsection{Model Watermarking}
There are plenty of works exploring model watermarking in other domains, such as computer vision \cite{uchida2017embedding,chen2019deepmarks,wang2020watermarking,zhang2018protecting,adi2018turning,guo2018watermarking,zhu2020secure,le2020adversarial}, graph learning \cite{zhao2021watermarking,xu2021watermarking}, federated learning \cite{shao2022fedtracker,liang2023fedcip,atli2020waffle,li2022fedipr,liu2021secure,yang2022watermarking,chen2023fedright}, etc. They can be divided into white-box and black-box techniques. White-box methods assume defenders can access the model parameters of a suspicious model, whereas black-box methods assume defenders can only query the suspicious model and observe the output.

White-box methods usually embed the watermark into model parameters. In computer vision, Uchida et al. \cite{uchida2017embedding} encrypts the watermark into the model parameters by multiplying the parameters with a key matrix. Successive works \cite{chen2019deepmarks,wang2020watermarking} mainly follow this design. In federated learning, the goal of model watermarking has shifted to identifying the malicious client, but the basic idea is still to embed the watermark into model parameters \cite{shao2022fedtracker,liang2023fedcip}.

Black-box methods assign a selected label to predefined backdoor patterns such as out-of-distribution data \cite{zhang2018protecting,adi2018turning,zhu2020secure} or dedicated noise \cite{guo2018watermarking,le2020adversarial}, and use such data to train the target model, so that when querying the model with these patterns, the model will predict the selected label. In graph learning, several works \cite{zhao2021watermarking,xu2021watermarking} have proposed to use ER random graphs as the watermark and assign predefined labels to the nodes. There are also a large number of works exploring black-box model watermarking on federated learning \cite{atli2020waffle,li2022fedipr,liu2021secure,yang2022watermarking,chen2023fedright}. Their methods basically follow the ideas in computer vision, but innovatively adopted to the federated scenario.

\subsection{Watermarking Against Distillation}
Yang et al. \cite{yang2019effectiveness} found that the above OOD watermarks can be easily removed by a model distillation. They propose to train an ingrainer model on the watermarks. Then it is fixed as a teacher to train the target model. Successive works mainly focus on generating in-distribution and real-like watermark samples. Li et al. \cite{li2019prove} propose to generate real-like watermark samples using a GAN. Namba et al. \cite{namba2019robust} propose to generate watermarks by relabeling benign samples as target watermark labels. Jia et al. \cite{jia2021entangled} propose to add triggers to in-distribution images. Szyller et al. \cite{szyller2021dawn} propose to alter the output label of the target model. The selection of altering is conducted by hashing. This method inevitably degrades the utility of the target model. All these methods cannot be simply transferred to the recommender systems due to the natural gap between the data format as well as the gap between the classification task and the recommendation task.

%% file: methodology.tex
\section{Methodology}
In this section, we introduce our method, Autoregressive Out-of-distribution Watermarking (AOW). We first present the problem definition, then describe the challenges to be addressed. Next, we discuss the possible choices when designing the watermarking technique and the advantages and disadvantages of each of them. At last, we introduce the details of AOW.

\begin{figure*}[t]
\centering
    \includegraphics[width=1.0\textwidth]{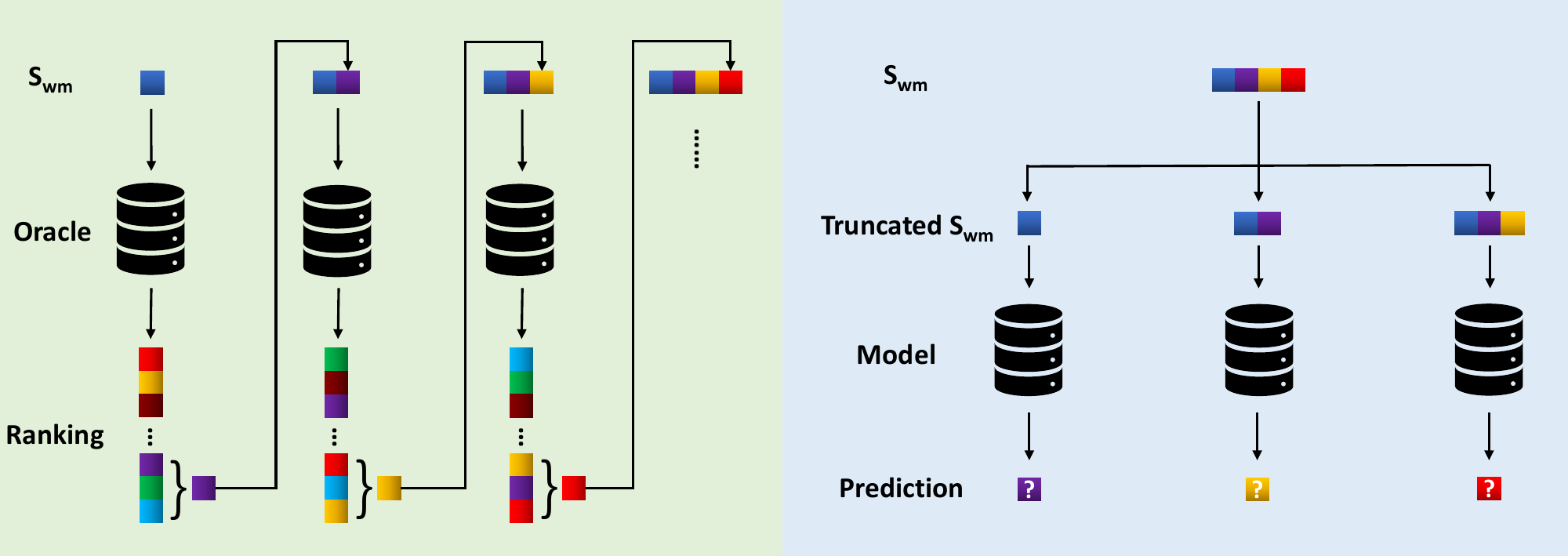}
\vspace{-2em}
\caption{An illustration of AOW. (Left) Generation process of the watermark sequence $S_{wm}$. An initial item (blue) is used to query the oracle model to obtain a ranking list, and a random item (purple) ranked at the bottom is selected as the next item. Then the new $S_{wm}$ that contains two items is used to query the oracle to obtain the third item (yellow). This process is repeated autoregressively until $S_{wm}$ reaches a predefined length. (Right) The evaluation process of AOW. The watermark sequence $S_{wm}$ is truncated into several subsequences. The model needs to predict the next item for each truncated sequence. The validity of the watermark is evaluated by the ranking position of the next item.}
\label{fig:method}
\vspace{-1em}
\end{figure*}

\subsection{Problem Definition}
Given a set of users $\mathcal{U}$ and a set of items $\mathcal{I}$, each user is associated with a sequence of interacted items, $S_{u}=\{i^{u}_{1},i^{u}_{2},...\}$, where $i^{u}_{j}$ denotes the $j$-th item interacted by user $u$. The set of all such sequences is $\mathcal{S}$, and is used to train a recommender model $f$. We call this model as the oracle model. Our goal is to design an additional sequence $S_{wm}$ to serve as the watermark. $S_{wm}$ and $\mathcal{S}$ will be used jointly to train a watermarked model $f_{wm}$ that can memorize the watermark sequence $S_{wm}$ while maintaining a good utility on $\mathcal{S}$.

\subsection{Challenges}
There are several constraints we need to consider before designing the watermark sequence.
\begin{itemize}
    \item \textbf{Model utility}. After injecting the watermark sequence into the training set, the utility of the model should be minimally affected. The performance of the model should be preserved as much as possible.
    \item \textbf{Watermark validity}. The confidence of the watermark should be high in a watermarked model. Meanwhile, a model trained without the watermark should have a low confidence on the watermark pattern.
    \item \textbf{Robustness}. The watermark should be robust against removal attacks such as distillation and fine-tuning. We should be able to detect the watermark with a high confidence under such removal attacks.
\end{itemize}

\subsection{Discussion}
In this section, we discuss what a good watermark design should be like and how AOW can achieve all of them.

\paragraph{\textbf{Black-box vs. White-box}} White-box watermarks assume that the model owner has access to a suspicious model's architecture and parameters. In this scenario, the owner can simply encode the watermark in the model parameters, which is already a well-studied technique in previous works \cite{uchida2017embedding,chen2019deepmarks,wang2020watermarking} and can be adopted to recommender systems directly. However, it is not always feasible to access the parameters of a suspicious model. So we focus on black-box watermarking where we can only query the suspicious model and observe its output.

\paragraph{\textbf{Out-of-distribution vs. In-distribution}} Previous black-box watermark designs in computer vision domain can be categorized into out-of-distribution (OOD) watermarks \cite{zhang2018protecting,adi2018turning,guo2018watermarking} and in-distribution (ID) watermarks \cite{yang2019effectiveness,li2019prove,namba2019robust,szyller2021dawn}. OOD watermarks are embedded in samples that do not follow the same distribution as the original dataset, while ID watermarks are embedded in samples that follow the original distribution. OOD watermarks do little harm to the model utility, but are not robust against removal attacks. On the contrary, ID watermarks are hard to be removed, but they inevitably degrade the model utility. Since recommender systems are highly correlated to commercial profits, utility degradation is unacceptable in many cases. Therefore, we choose to design OOD watermarks. However, as we will show in the experiment section, AOW is also effective in resisting removal attacks.

\paragraph{\textbf{Fake items}} One possible solution to construct OOD watermarks is to create fake items that do not exist in the original dataset. This is almost guaranteed to be OOD because the model has no information regarding the fake items. For example, if a query sequence contains a fake item, then we can force the model to output a designated label or ranking list. However, this approach is not always feasible and effective in real situations. On the one hand, some model distillation attacks assume the attacker can acquire in-distribution data, which, however, does not include the fake item. Therefore, it is difficult for the distilled model to inherit the watermark. On the other hand, the service provider may need to assign a real item as the fake item to avoid it being recognized or detected. However, due to the complicated environments and demands in different recommendation scenarios, this has to be done with domain-specific expert knowledge. Therefore, in order to design a more universal method that can be applied to most recommender systems, we do not use fake items.

\paragraph{\textbf{The choice of the watermark pattern}} Since fake items are not used in our method, we have to use existing items to form a special input-output mapping as the watermark. The form of this mapping can be diverse. For example, for the input, the watermark trigger can be either an entire sequence or a subsequence with filler items. To be specific, one can force the model to produce a desired output given a sequence with all the items fixed, or given a sequence where some of the items are predetermined and other items are randomly selected. However, if using a subsequence, it is at the risk of losing model utility, because it is difficult to guarantee that the complete sequence is an OOD sequence, since the choices of filler items might result in the whole sequence shifting towards an in-distribution sequence, which will disrupt the training of benign sequences. For the output, we can check the existence of certain items in the top-k ranking, or we can check the existence of a special pattern with some predefined items following a fixed ordering. However, the latter that checks the ordering can be easily removed by doing global or local shuffling on the output top-k ranking list even if the attacker is not aware of the watermark. Therefore, we choose to use an entire sequence as the watermark, and ask the model to predict the next item given all the previous items.

\subsection{Autoregressive Out-of-distribution Watermarking (AOW)}
\label{sec:AOW}
To this end, we propose to generate an entire sequence $S_{wm}=\{i^{wm}_{1},i^{wm}_{2},...,i^{wm}_{n}\}$ as the watermark. The length $n$ is a hyperparameter to be specified by the model owner. The model should be able to memorize this sequence by predicting the $j$-th item $i^{wm}_{j}$ when given the previous $j-1$ items $\{i^{wm}_{1},i^{wm}_{2},...,i^{wm}_{j-1}\}$. We first train an oracle model using the original dataset, then generate $S_{wm}$ autoregressively. An illustration of AOW is shown in \autoref{fig:method}. Specifically, we first select an initial item $i^{wm}_{1}$ (the selection of the initial item will be discussed in the experiment section), then query the oracle model with this single item to get scores of all items. We then rank all items according to the scores and focus on a certain number of items with the lowest scores (let's say bottom-M in contrast to top-k, where the setting of M will be discussed later on). We draw one item randomly from the bottom-M items, and append it after the initial item to expand the sequence. For model owners who want to design their unique watermark pattern, they can pick one particular item from the bottom-M items by themselves. This may also serve as an effectiveness fingerprinting technique in federated recommendation \cite{yuan2023manipulating,yuan2023interaction,qu2024towards,yuan2023hide,zhang2023comprehensive} to detect malicious clients who are responsible for the model leakage, which we leave for future work. Now the sequence contains two items, and we query the oracle model again with it and repeat the above process to get the third item. We stop when the sequence reaches a predefined length. This sequence is defined as the watermark sequence $S_{wm}$. It will be used together with the original training set to train a new model $f_{wm}$, and this model will be able to memorize $S_{wm}$ while maintaining its utility. For evaluation, we generate $n-1$ truncated sequences from $S_{wm}$. For example, $S^{1}_{wm}=\{i^{wm}_{1}\}$, $S^{2}_{wm}=\{i^{wm}_{1},i^{wm}_{2}\}$, ..., $S^{n-1}_{wm}=\{i^{wm}_{1},i^{wm}_{2},...,i^{wm}_{n-1}\}$. The model needs to predict the next item given these truncated sequences. A well-memorized watermark should have the ground-truth next item ranked as high as possible. This design has several advantages.
\begin{itemize}
    \item $S_{wm}$ is highly likely an OOD sequence. Recommender systems tend to give high scores to in-distribution sequences when making predictions. Therefore, if an item has a low score in the prediction, it suggests that this item does not follow the distribution of the training dataset. Therefore, we can generate an OOD sequence by autoregressively selecting such items.
    \item The oracle model is almost guaranteed to have poor performance on $S_{wm}$. As long as the set of the bottom-M items does not overlap with the set of the top-k items, the oracle model will always have zero recall and NDCG on $S_{wm}$ in terms of top-k evaluations. For example, assume there are 1000 items in total, and we set $M=100$ and $k=20$. For a truncated $S_{wm}$, for example $\{i^{wm}_{1},i^{wm}_{2},i^{wm}_{3}\}$, the next item $i^{wm}_{4}$ must appear in the bottom-100 since this is how we selected it. So any evaluation metric that only considers top-20 will never find $i^{wm}_{4}$ there, because it is ranked outside 900. However, if there are 110 items in total, and we still set M=100 and k=20, then the top-20 and bottom-100 will overlap, and $i^{wm}_{4}$ may appear in both. In practice, a real-world dataset usually contains at least hundreds or thousands of items, so it should always be feasible to set appropriate M and k. In fact, we have conducted experiments to try to detect the watermark from the oracle model, and unsurprisingly, all return zero recall and NDCG.
    \item Simple tricks would not affect the watermark. As discussed before, some tricks such as shuffling can disturb the watermark if it is encoded as a special ordering even if the attacker has no knowledge about the watermark. However, for AOW, the only way for removing the watermark is to demote the target item from the top-k ranking. Having no idea of what the target item is, the attacker can only randomly choose some items and change their rankings. This will result in a significant utility degradation.
\end{itemize}

%% file: experiments.tex
\section{Experiments}
In this section, we aim to answer the following research questions regarding the performance of AOW:
\begin{itemize}
    \item \textbf{RQ1}: How is the validity of the watermark and the utility of the model? Can we extract the watermark with a high confidence while preserving the model utility?
    \item \textbf{RQ2}: How does AOW perform against distillation and fine-tuning?
    \item \textbf{RQ3}: How do the choices of hyperparameters, including the watermark sequence length $n$, the initial item, the watermark-to-data ratio, and the parameter M in bottom-M item selection range influence its performance?
\end{itemize}
We will first introduce our experiment settings, then present results to answer these questions.

\begin{table}[t]
\caption{Dataset Statistics}
\vspace{-1.5em}
\begin{center}
 \begin{tabular}{|c|c|c|c|c|} 
 \hline
 Dataset & \# Users & \# Items & Avg. length & Density \\
 \hline
 ML-1M & 6,040 & 3,416 & 165.5 & 4.84\% \\
 \hline
 ML-20M & 138,493 & 18,345 & 144.3 & 0.79\% \\
 \hline
 Steam & 334,542 & 13,046 & 12.6 & 0.10\% \\
 \hline
 Beauty & 40,226 & 54,542 & 8.8 & 0.02\% \\
 \hline
\end{tabular}
\label{table:datasets}
\end{center}
\vspace{-1.5em}
\end{table}

\begin{table*}[t]
\caption{Main experiment results in percentage for each dataset. We set the initial item as the cold item, and fix the length of the watermark as 5. The base model is Bert4Rec. (1) The 2nd column shows the validity of the watermark on the target model with Recall@1. (2) The 3rd-5th columns show the utility (Recall@10) of the target model under no protection (Oracle), trained with GRO, and trained with AOW respectively. (3) The 6th column shows the watermark validity of the model after distillation using the same model architecture. (4) The 7th column shows the watermark validity (Recall@10) after fine-tuning with 1\% fine-tuning sequences.}
\vspace{-1.5em}
\begin{center}
 \begin{tabular}{|c|c|c|c|c|c|c|} 
 \hline
  \multirow{2}{*}{Dataset} & \multirow{2}{*}{Validity (R@1)} & \multicolumn{3}{c|}{Utility (R@10)} & \multirow{2}{*}{Validity After Distillation (R@10)} & \multirow{2}{*}{Validity After Fine-tuning (R@10)}\\
 \cline{3-5}
   & & Oracle & GRO & AOW  & &  \\
  \hline
 ML-1M & 100.00 & 20.16 & 18.08 & 19.69 & 100.00 & 100.00 \\
  \hline
   ML-20M & 100.00 & 14.96 & 13.95 & 14.35 & 100.00 & 100.00 \\
 \hline
 Steam & 100.00 & 19.93 & 19.87 & 19.90 & 100.00 & 100.00 \\
  \hline
  Beauty & 100.00 & 2.96 & 2.85 & 2.95 & 75.27 & 100.00\\
 \hline
\end{tabular}
\label{table:main}
\end{center}
\vspace{-1em}
\end{table*}

\begin{table}[t]
\caption{Recall@1 in percentage of the watermark at different watermark lengths $n$ on Bert4Rec. The method $\textit{cold}$ denotes using the item with the least number of interactions as the initial item, while $\textit{pop}$ denotes using the most popular item which has the most number of interactions as the initial item.}
\vspace{-1.5em}
\begin{center}
 \begin{tabular}{|c|c|c|c|c|c|} 
 \hline
  \multirow{2}{*}{Dataset} & \multirow{2}{*}{Method} & \multicolumn{4}{c|}{$n$} \\
 \cline{3-6}
   & & 2 & 5 & 10 & 20 \\
  \hline
 \multirow{2}{*}{ML-1M} & cold & 100.00 & 100.00 & 100.00 & 100.00 \\
   \cline{2-2}
 & pop & 100.00 & 100.00 & 100.00 & 100.00 \\
  \hline
   \multirow{2}{*}{ML-20M} & cold & 100.00 & 100.00 & 100.00 & 100.00 \\
   \cline{2-2}
 & pop & 100.00 & 100.00 & 100.00 & 100.00 \\
 \hline
 \multirow{2}{*}{Steam} & cold & 100.00 & 100.00 & 100.00 & 91.01 \\
   \cline{2-2}
 & pop & 100.00 & 100.00 & 88.18 & 85.79 \\
  \hline
  \multirow{2}{*}{Beauty} & cold & 100.00 & 100.00 & 100.00 & 100.00  \\
   \cline{2-2}
 & pop & 100.00 & 100.00 & 100.00 & 93.89 \\
 \hline
\end{tabular}
\label{table:validity}
\end{center}
\vspace{-1em}
\end{table}


\begin{table*}[t]
\caption{Model utility in percentage on Bert4Rec. For AOW, we set watermark length to 5 and use the cold item as the initial item. Other hyperparameter settings yield similar results.}
\vspace{-1.5em}
\begin{center}
 \begin{tabular}{|c|c|c|c|c|c|c|c|c|c|c|c|} 
 \hline
  Dataset & Method & R@1 & R@5 & R@10 & R@20 & R@100 & N@1 & N@5 & N@10 & N@20 & N@100 \\
    \hline
   \multirow{3}{*}{ML-1M}& Oracle & 3.65 & 12.90 & 20.16 & 30.73 & 60.60 & 3.65 & 8.26 & 10.59 & 13.25 & 18.72 \\
 \cline{2-2}
 & GRO & 2.58 & 10.20 & 18.08 & 28.96 & 58.83 & 2.58 & 6.81 & 9.53 & 12.48 & 16.97 \\
  \cline{2-2}
 & AOW & 3.50 & 12.64 & 19.69 & 30.40 & 60.31 & 3.50 & 8.06 & 10.34 & 13.03 & 18.51 \\
  \hline
   \multirow{3}{*}{ML-20M} & Oracle & 2.80 & 9.26 & 14.96 & 22.58 & 47.05 & 2.80 & 6.02 & 7.85 & 9.77 & 14.20 \\
 \cline{2-2}
 & GRO & 2.26 & 7.95 & 13.95 & 21.63 & 45.79 & 2.26 & 5.12 & 7.23 & 9.02 & 13.62 \\
 \cline{2-2}
 & AOW & 2.70 & 8.95 & 14.35 & 21.96 & 46.61 & 2.70 & 5.81 & 7.55 & 9.46 & 13.93 \\
 \hline
  \multirow{3}{*}{Steam} & Oracle & 12.13 & 16.44 & 19.93 & 24.94 & 43.72 & 12.13 & 14.31 & 15.43 & 16.69 & 20.06 \\
 \cline{2-2}
  & GRO & 11.99 & 16.28 & 19.87 & 24.69 & 42.92 & 11.99 & 14.10 & 15.39 & 16.58 & 19.80 \\
 \cline{2-2}
  & AOW & 12.09 & 16.36 & 19.90 & 24.83 & 43.24 & 12.09 & 14.24 & 15.40 & 16.62 & 19.90 \\
  \hline
  \multirow{3}{*}{Beauty} & Oracle & 0.42 & 1.70 & 2.96 & 4.77 & 11.74 & 0.42 & 1.06 & 1.46 & 1.92 & 3.16 \\
 \cline{2-2}
  & GRO & 0.45 & 1.69 & 2.85 & 4.48 & 11.06 & 0.45 & 1.07 & 1.44 & 1.85 & 3.03 \\
 \cline{2-2}
  & AOW & 0.44 & 1.75 & 2.95 & 4.69 & 11.45 & 0.44 & 1.10 & 1.48 & 1.91 & 3.13 \\
 \hline
\end{tabular}
\label{table:model utility}
\end{center}
\vspace{-1em}
\end{table*}

\begin{table*}[t]
\caption{Watermark validity in percentage on the distilled Bert4Rec model.}
\vspace{-1.5em}
\begin{center}
 \begin{tabular}{|c|c|c|c|c|c|c|c|c|c|c|c|c|} 
 \hline
 Dataset & Method & $n$ & R@1 & R@5 & R@10 & R@20 & R@100 & N@1 & N@5 & N@10 & N@20 & N@100 \\
    \hline
   \multirow{8}{*}{ML-1M} & \multirow{4}{*}{cold} & 2 & 100.00 & 100.00 & 100.00 & 100.00 & 100.00 & 100.00 & 100.00 & 100.00 & 100.00 & 100.00 \\
     \cline{3-3}
 &  & 5 & 26.09 & 100.00 & 100.00 & 100.00 & 100.00 & 26.09 & 69.69 & 69.69 & 69.69 & 69.69 \\
  \cline{3-3}
 &  & 10 & 35.22 & 100.00 & 100.00 & 100.00 & 100.00 & 35.22 & 70.08 & 70.08 & 70.08 & 70.08 \\
  \cline{3-3}
  &  & 20 & 0.00 & 6.51 & 11.21 & 26.13 & 57.57 & 0.00 & 2.52 & 3.93 & 7.81 & 13.44 \\
 \cline{2-13}
 & \multirow{4}{*}{pop} & 2 & 100.00 & 100.00 & 100.00 & 100.00 & 100.00 & 100.00 & 100.00 & 100.00 & 100.00 & 100.00 \\
  \cline{3-3}
   & & 5 & 72.16 & 100.00 & 100.00 & 100.00 & 100.00 & 72.16 & 86.08 & 86.08 & 86.08 & 86.08 \\
 \cline{3-3}
  & & 10 & 43.84 & 100.00 & 100.00 & 100.00 & 100.00 & 43.84 & 77.81 & 77.81 & 77.81 & 77.81 \\
  \cline{3-3}
  & & 20 & 10.46 & 28.19 & 59.81 & 90.76 & 100.00 & 10.46 & 17.32 & 27.03 & 34.85 & 36.78 \\
\hline
   \multirow{8}{*}{ML-20M} & \multirow{4}{*}{cold} & 2 & 100.00 & 100.00 & 100.00 & 100.00 & 100.00 & 100.00 & 100.00 & 100.00 & 100.00 & 100.00 \\
     \cline{3-3}
 &  & 5 & 73.95 & 100.00 & 100.00 & 100.00 & 100.00 & 73.95 & 90.39 & 90.39 & 90.39 & 90.39 \\
  \cline{3-3}
 &  & 10 & 100.00 & 100.00 & 100.00 & 100.00 & 100.00 & 100.00 & 100.00 & 100.00 & 100.00 & 100.00 \\
  \cline{3-3}
  &  & 20 & 91.58 & 100.00 & 100.00 & 100.00 & 100.00 & 91.58 & 96.89 & 96.89 & 96.89 & 96.89 \\
 \cline{2-13}
 & \multirow{4}{*}{pop} & 2 & 100.00 & 100.00 & 100.00 & 100.00 & 100.00 & 100.00 & 100.00 & 100.00 & 100.00 & 100.00 \\
  \cline{3-3}
 & & 5 & 100.00 & 100.00 & 100.00 & 100.00 & 100.00 & 100.00 & 100.00 & 100.00 & 100.00 & 100.00 \\
 \cline{3-3}
 & & 10 & 88.96 & 100.00 & 100.00 & 100.00 & 100.00 & 88.96 & 95.93 & 95.93 & 95.93 & 95.93 \\
  \cline{3-3}
 & & 20 & 90.91 & 95.83 & 95.83 & 100.00 & 100.00 & 90.91 & 93.37 & 93.37 & 94.39 & 94.39 \\
\hline
   \multirow{8}{*}{Steam} & \multirow{4}{*}{cold} & 2 & 100.00 & 100.00 & 100.00 & 100.00 & 100.00 & 100.00 & 100.00 & 100.00 & 100.00 & 100.00 \\
     \cline{3-3}
 &  & 5 & 24.83 & 100.00 & 100.00 & 100.00 & 100.00 & 24.83 & 62.71 & 62.71 & 62.71 & 62.71 \\
  \cline{3-3}
 &  & 10 & 22.76 & 58.91 & 90.38 & 90.38 & 100.00 & 22.76 & 37.76 & 48.33 & 48.33 & 49.83 \\
  \cline{3-3}
  &  & 20 & 0.00 & 33.72 & 69.07 & 89.27 & 94.37 & 0.00 & 15.94 & 27.24 & 32.56 & 33.56 \\
 \cline{2-13}
 & \multirow{4}{*}{pop} & 2 & 100.00 & 100.00 & 100.00 & 100.00 & 100.00 & 100.00 & 100.00 & 100.00 & 100.00 & 100.00 \\
  \cline{3-3}
   & & 5 & 25.39 & 75.38 & 75.38 & 75.38 & 100.00 & 25.39 & 52.10 & 52.10 & 52.10 & 56.76 \\
 \cline{3-3}
  & & 10 & 0.00 & 11.24 & 66.72 & 77.68 & 77.68 & 0.00 & 7.09 & 24.95 & 27.91 & 27.91 \\
  \cline{3-3}
  & & 20 & 5.70 & 30.84 & 58.04 & 90.20 & 94.76 & 5.70 & 17.90 & 26.89 & 35.14 & 36.11 \\
\hline
\multirow{8}{*}{Beauty} & \multirow{4}{*}{cold} & 2 & 100.00 & 100.00 & 100.00 & 100.00 & 100.00 & 100.00 & 100.00 & 100.00 & 100.00 & 100.00 \\
  \cline{3-3}
 &  & 5 & 25.02 & 75.27 & 75.27 & 75.27 & 75.27 & 25.02 & 52.03 & 52.03 & 52.03 & 52.03 \\
  \cline{3-3}
 &  & 10 & 0.00 & 0.00 & 0.00 & 11.40 & 11.40 & 0.00 & 0.00 & 0.00 & 2.99 & 2.99 \\
  \cline{3-3}
  &  & 20 & 0.00 & 0.00 & 0.00 & 0.00 & 5.12 & 0.00 & 0.00 & 0.00 & 0.00 & 0.82 \\
  \cline{2-13}
 & \multirow{4}{*}{pop} & 2 & 100.00 & 100.00 & 100.00 & 100.00 & 100.00 & 100.00 & 100.00 & 100.00 & 100.00 & 100.00 \\
  \cline{3-3}
   & & 5 & 74.89 & 100.00 & 100.00 & 100.00 & 100.00 & 74.89 & 90.73 & 90.73 & 90.73 & 90.73 \\
 \cline{3-3}
  & & 10 & 21.60 & 55.69 & 55.69 & 55.69 & 68.16 & 21.61 & 37.99 & 37.99 & 37.99 & 39.89 \\
  \cline{3-3}
  & & 20 & 9.23 & 26.34 & 30.99 & 30.99 & 63.33 & 9.23 & 18.55 & 19.95 & 19.95 & 25.75 \\
    \hline
\end{tabular}
\label{table:distillation}
\end{center}
\vspace{-1.5em}
\end{table*}

\begin{table*}[t]
\caption{Watermark validity in percentage on the fine-tuned Bert4Rec model on ML-20M. Data size refers to the ratio of number of fine-tuning sequences to the number of training sequences.}
\vspace{-1.5em}
\begin{center}
 \begin{tabular}{|c|c|c|c|c|c|c|c|c|c|c|c|c|} 
 \hline
 Data size & Method & $n$ & R@1 & R@5 & R@10 & R@20 & R@100 & N@1 & N@5 & N@10 & N@20 & N@100 \\
    \hline
   \multirow{8}{*}{1\%} & \multirow{4}{*}{cold} & 2 & 100.00 & 100.00 & 100.00 & 100.00 & 100.00 & 100.00 & 100.00 & 100.00 & 100.00 & 100.00 \\
     \cline{3-3}
 & & 5 & 100.00 & 100.00 & 100.00 & 100.00 & 100.00 & 100.00 & 100.00 & 100.00 & 100.00 & 100.00 \\
  \cline{3-3}
 &  & 10 & 87.42 & 100.00 & 100.00 & 100.00 & 100.00 & 87.42 & 95.36 & 95.36 & 95.36 & 95.36 \\
  \cline{3-3}
  &  & 20 & 84.13 & 100.00 & 100.00 & 100.00 & 100.00 & 84.13 & 94.14 & 94.14 & 94.14 & 94.14 \\
 \cline{2-13}
 & \multirow{4}{*}{pop} & 2 & 100.00 & 100.00 & 100.00 & 100.00 & 100.00 & 100.00 & 100.00 & 100.00 & 100.00 & 100.00 \\
  \cline{3-3}
   & & 5 & 100.00 & 100.00 & 100.00 & 100.00 & 100.00 & 100.00 & 100.00 & 100.00 & 100.00 & 100.00 \\
 \cline{3-3}
  & & 10 & 100.00 & 100.00 & 100.00 & 100.00 & 100.00 & 100.00 & 100.00 & 100.00 & 100.00 & 100.00 \\
  \cline{3-3}
  & & 20 & 90.33 & 95.40 & 95.40 & 95.40 & 95.40 & 90.33 & 93.53 & 93.53 & 93.53 & 93.53 \\
    \hline
    \multirow{8}{*}{5\%} & \multirow{4}{*}{cold} & 2 & 100.00 & 100.00 & 100.00 & 100.00 & 100.00 & 100.00 & 100.00 & 100.00 & 100.00 & 100.00 \\
      \cline{3-3}
 &  & 5 & 50.36 & 74.28 & 74.28 & 74.28 & 100.00 & 50.36 & 65.45 & 65.45 & 65.45 & 70.47 \\
  \cline{3-3}
 &  & 10 & 19.87 & 54.97 & 77.50 & 77.50 & 77.50 & 19.87 & 38.07 & 45.02 & 45.02 & 45.02 \\
  \cline{3-3}
  &  & 20 & 5.90 & 25.98 & 25.98 & 25.98 & 45.82 & 5.90 & 16.96 & 16.96 & 16.96 & 20.64 \\
 \cline{2-13}
 & \multirow{4}{*}{pop} & 2 & 100.00 & 100.00 & 100.00 & 100.00 & 100.00 & 100.00 & 100.00 & 100.00 & 100.00 & 100.00 \\
  \cline{3-3}
   & & 5 & 76.97 & 76.97 & 76.97 & 76.97 & 76.97 & 76.97 & 76.97 & 76.97 & 76.97 & 76.97 \\
 \cline{3-3}
  & & 10 & 66.93 & 89.50 & 89.50 & 89.50 & 89.50 & 66.93 & 81.17 & 81.17 & 81.17 & 81.17 \\
  \cline{3-3}
  & & 20 & 0.00 & 5.70 & 11.48 & 21.20 & 21.20 & 0.00 & 3.60 & 5.42 & 8.02 & 8.02 \\
    \hline
   \multirow{8}{*}{10\%} & \multirow{4}{*}{cold} & 2 & 100.00 & 100.00 & 100.00 & 100.00 & 100.00 & 100.00 & 100.00 & 100.00 & 100.00 & 100.00 \\
     \cline{3-3}
 &  & 5 & 100.00 & 100.00 & 100.00 & 100.00 & 100.00 & 100.00 & 100.00 & 100.00 & 100.00 & 100.00 \\
  \cline{3-3}
 &  & 10 & 90.15 & 100.00 & 100.00 & 100.00 & 100.00 & 90.15 & 96.37 & 96.37 & 96.37 & 96.37 \\
  \cline{3-3}
  &  & 20 & 68.46 & 100.00 & 100.00 & 100.00 & 100.00 & 68.46 & 84.14 & 84.14 & 84.14 & 84.14 \\
 \cline{2-13}
 & \multirow{4}{*}{pop} & 2 & 100.00 & 100.00 & 100.00 & 100.00 & 100.00 & 100.00 & 100.00 & 100.00 & 100.00 & 100.00 \\
  \cline{3-3}
   & & 5 & 100.00 & 100.00 & 100.00 & 100.00 & 100.00 & 100.00 & 100.00 & 100.00 & 100.00 & 100.00 \\
 \cline{3-3}
  & & 10 & 100.00 & 100.00 & 100.00 & 100.00 & 100.00 & 100.00 & 100.00 & 100.00 & 100.00 & 100.00 \\
  \cline{3-3}
  & & 20 & 26.78 & 36.99 & 53.22 & 68.58 & 73.55 & 26.78 & 31.88 & 37.31 & 41.07 & 41.84 \\
    \hline
\multirow{8}{*}{20\%} & \multirow{4}{*}{cold} & 2 & 0.00 & 0.00 & 0.00 & 0.00 & 100.00 & 0.00 & 0.00 & 0.00 & 0.00 & 15.64 \\
  \cline{3-3}
 &  & 5 & 0.00 & 22.96 & 22.96 & 22.96 & 22.96 & 0.00 & 9.89 & 9.89 & 9.89 & 9.89 \\
  \cline{3-3}
 &  & 10 & 12.29 & 12.29 & 12.29 & 12.29 & 21.23 & 12.29 & 12.29 & 12.29 & 12.29 & 13.72 \\
  \cline{3-3}
  &  & 20 & 0.00 & 0.00 & 0.00 & 0.00 & 10.07 & 0.00 & 0.00 & 0.00 & 0.00 & 1.65 \\
 \cline{2-13}
 & \multirow{4}{*}{pop} & 2 & 0.00 & 0.00 & 0.00 & 0.00 & 0.00 & 0.00 & 0.00 & 0.00 & 0.00 & 0.00 \\
  \cline{3-3}
   & & 5 & 0.00 & 25.54 & 25.54 & 25.54 & 25.54 & 0.00 & 12.77 & 12.77 & 12.77 & 12.77 \\
 \cline{3-3}
  & & 10 & 0.00 & 0.00 & 0.00 & 9.41 & 9.41 & 0.00 & 0.00 & 0.00 & 2.14 & 2.14 \\
  \cline{3-3}
  & & 20 & 15.55 & 29.97 & 35.13 & 45.39 & 55.87 & 15.55 & 22.87 & 24.71 & 27.52 & 29.60 \\
    \hline
\end{tabular}
\label{table:fine-tuning}
\end{center}
\vspace{-1.5em}
\end{table*}

\subsection{Datasets and Evaluation Metrics}
We use four publicly available datasets, namely MovieLens-1M, MovieLens-20M\footnote{\url{https://grouplens.org/datasets/movielens/}}, Amazon-Beauty\footnote{\url{https://cseweb.ucsd.edu/~jmcauley/datasets/amazon/links.html}}, and Steam\footnote{\url{https://cseweb.ucsd.edu/~jmcauley/datasets.html\#steam_data}}. The statistics of these datasets are summarized in \autoref{table:datasets}. Each user is associated with one interaction sequence. We use leave-one-out evaluation, where we leave the last item of each sequence as the test set, and the second last item as the validation set. For evaluation metrics, we use Recall@k and Normalized Discounted Cumulative Gain@k (NDCG@k) to measure both the utility of the model and the validity of the watermark.

\subsection{Baseline and Model}
One advantage of model watermarking is the ability to maintain the utility. We show it by comparing AOW with GRO \cite{zhang2024defense}, an defense method against model extraction attacks on recommender systems, which can also be used to protect the model intellectual property. GRO defines a swap loss and jointly trains with the original task. Although it can bring utility degradation to the attacker's model, it will also harm the utility of the target model. We make an comparison between the utility of the target model under AOW and GRO, and demonstrate the effectiveness of AOW in preserving the utility. Following their work, we choose Bert4Rec \cite{sun2019bert4rec} as the backbone model, and evaluate model performance by ranking all the items in the dataset. However, AOW is a model-agnostic approach that can be applied to arbitrary sequential models. We have also conducted experiments on another model SASRec \cite{kang2018self}, and the results and conclusions are similar with Bert4Rec. Therefore, we only present the results on Bert4Rec for simplicity and consistency.

\subsection{Settings}
For AOW, we by default set the hyperparameter M for bottom-M as 100. An analysis of different choices of M will be shown in \autoref{sec:M}. We vary the length of the watermark sequence among 2, 5, 10, 20. The initial item is chosen between the \textit{cold} item and the \textit{pop} item, where \textit{cold} denotes the item with the least number of interactions, and \textit{pop} denotes the item with the most number of interactions. After the watermark sequence $S_{wm}$ has been generated, we set a predefined watermark-to-data ratio (WDR) to determine the weight of $S_{wm}$ during training. We can duplicate $S_{wm}$ for (WDR$\times |\mathcal{S}|$) times and append them after the training set. For example, ML-1M has 6,040 sequences (users). If we set WDR to 0.1, we will duplicate $S_{wm}$ for 604 times, resulting in a new training set of 6,644 sequences. Another way to achieve the same effect is to use a single $S_{wm}$ sequence but with a weighted loss function added to the loss on the original training set. Changing the weight parameter is equal to changing the WDR. Therefore, the additional training complexity introduced by AOW can be neglected. By default, we set WDR to 0.1. We provide a study on the impact of WDR in \autoref{sec:wdr}.

For evaluation, the validity of the watermark is measured by the average recall and NDCG on the truncated watermark sequences. Same as how we deal with the training set, we also append these truncated sequences to the validation set in order to select a model with good performance on both the original task and the watermark task. Since we did not observe a significant performance gap when altering the ratio between the regular validation set and the truncated watermark set, we by default set the weight of the truncated watermark set to be the same as the original validation set.

We follow the hyperparameter setting from Bert4Rec for each dataset. For GRO, we follow their official implementation and setting. For distillation, we use the black-box model extraction attack proposed by Yue et al. \cite{yue2021black} and follow their suggested setting to autoregressively generate 3,000 distillation sequences. For fine-tuning, we adopt the strategy by Yue et al. \cite{yue2021black} and generate some new sequences autoregressively through querying the oracle model. The number of generated sequences vary between 1\% to 20\% of the number of sequences in the original dataset. These generated sequences serve as an approximation of the in-distribution data that an attacker might have access to, but are different from the data used to train the oracle model. We use them to fine-tune the watermarked model. We fine-tune each model for the same number of epochs as during training.

\subsection{Main Results}
We first present the overall performance of AOW under a fixed hyperparameter setting in \autoref{table:main}. The initial item is the \textit{cold} item. The length of the watermark is 5. The second column shows the watermark validity (Recall@1), being 100\% across all datasets. This underscores the watermark's capacity to be effectively retained by the target model. The 3rd-5th columns show the utility (Recall@10) of the target model protected by different methods. It is clear that AOW consistently outperforms GRO in preserving the model utility. The 6th column and the 7th column show the watermark validity after model distillation and model fine-tuning, respectively. Both demonstrate high watermark robustness. Next, we present detailed studies on (1) watermark validity and model utility; (2) robustness against distillation and fine-tuning; (3) hyperparameter study.

\subsection{RQ1: Watermark Validity \& Model Utility}
\subsubsection{Watermark Validity}
We first test the validity of the watermark on a model trained with AOW. \autoref{table:validity} shows Recall@1 under two different choices for the initial item and four different watermark lengths. Specifically, we change the watermark length $n$ from 2 to 20, and change the initial item between the cold item and the popular item. We have three observations from the table. First, the watermark is successfully embedded in the model with Recall@1=1 for most cases. This suggests the effectiveness of our watermark. Second, a shorter watermark is easier for the model to memorize, as the recall decreases when the length increases. This is reasonable since longer sequences contain complex long-term dependencies which would be more challenging to memorize. Third, for the selection of the initial item, the cold one is better than the popular one when the watermark length is large (10 and 20). However, they both perform good with Recall@1=1 when the watermark length is small (2 and 5). Cold items are not as robust as popular items since there is less information about them. Thus it is easier for the model to memorize the watermark sequence which starts with the cold item.

\subsubsection{Model Utility}
We compare AOW with GRO to show the ability of AOW in preserving the utility of the target model. The results are listed in \autoref{table:model utility}. We report results from the case when the initial item is the cold item and the watermark length is 5, while the popular item as well as other watermark lengths yield similar results. The recall and NDCG of the target model trained with GRO are significantly lower than the oracle model, while AOW performs much better in preserving the utility, achieving a comparable performance with the oracle model, demonstrating the superior performance of AOW in preserving the model utility.

\subsection{RQ2: Against Distillation and Fine-tuning}
\subsubsection{Against Distillation}
We test how AOW performs against distillation. The recall and NDCG of the watermark validity are reported in \autoref{table:distillation}. 

First, as the watermark length $n$ increases, the watermark validity decreases. It can achieve 1.0 recall for all datasets with a small watermark length, but behaves poorly when the length increases. Although all the watermarks can be well memorized by the oracle as shown in \autoref{table:validity}, the distillation performance gap between short watermarks and long watermarks suggests that the short watermarks are easier for the model to memorize towards in-distribution data, whereas the long watermarks are still recognized as out-of-distribution ones, causing them to be easily removed by distillation.

Second, for the selection of the initial item, different datasets yield different results. On ML-1M and Beauty, the popular item outperforms the cold item, whereas on Steam, the cold item outperforms the popular item. On ML-20M, they behave on par. We assume this to be due to the distinct characteristics of these datasets, e.g., the amount of information that can be extracted and learned for each item, which will directly influence the robustness of the item embeddings and thus affect the validity of the injected watermark. If the embeddings are dense and robust, it will be hard to memorize the watermark that comes along as an OOD data. Therefore, the best initial item and the best watermark length should be subject to the specific model and dataset. But it should always be a good choice to use the cold initial item and a short watermark length, since cold items are always the least robust items in a recommender system.

\begin{figure}[t]
\centering
    \includegraphics[width=0.45\textwidth]{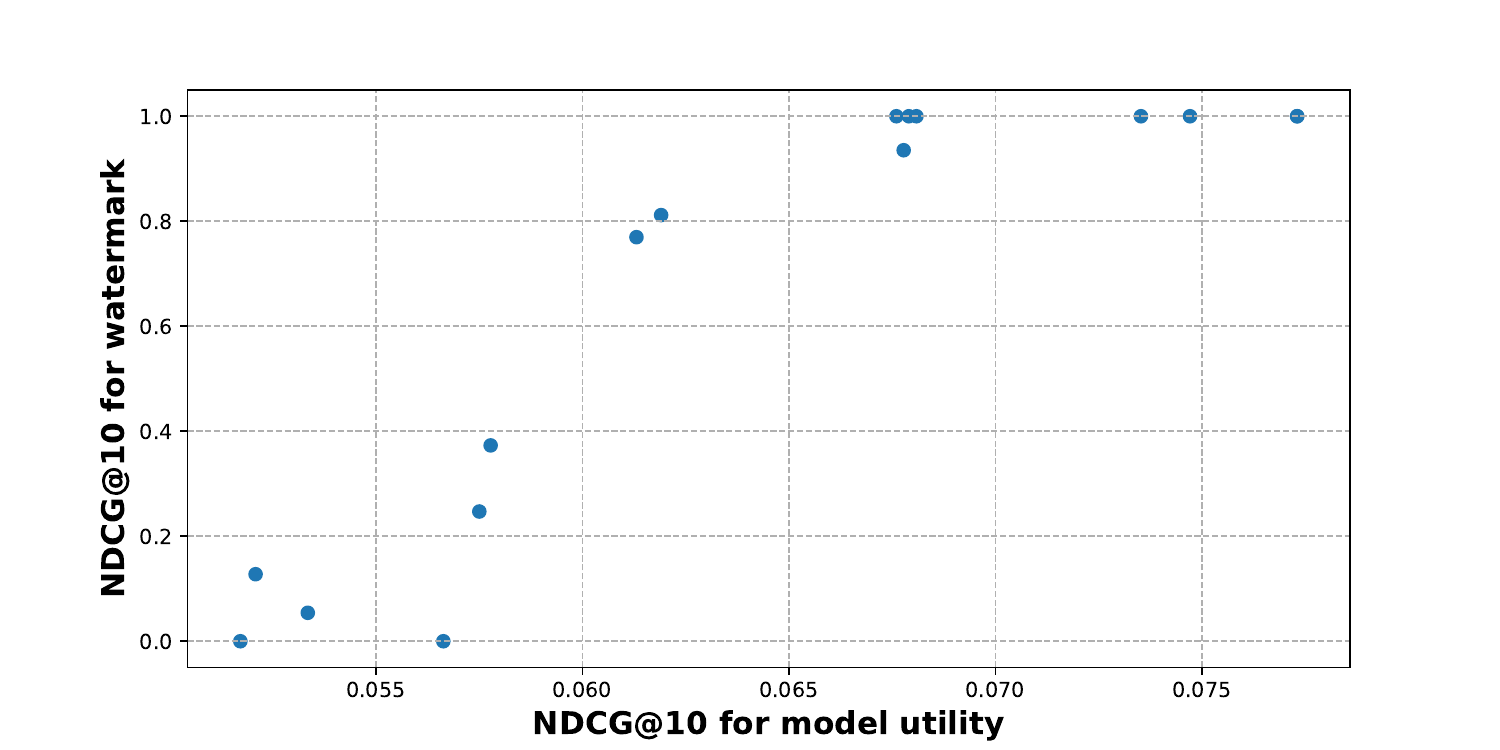}
\vspace{-1em}
\caption{Watermark validity vs. model utility after fine-tuning on ML-20M with the popular item as the initial item. Each point represents a model under different hyperparameters including the watermark length and the number of fine-tuning sequences. We report the NDCG@10 for the watermark validity in the y-axis, and the NDCG@10 for the model utility in the x-axis.}
\label{fig:validity vs utility}
\vspace{-2em}
\end{figure}

\subsubsection{Against Fine-tuning}
We then test how AOW performs against fine-tuning. The results are shown in \autoref{table:fine-tuning}. We report the recall and NDCG of the watermark on the fine-tuned model on ML-20M dataset. We alter the number of the generated fine-tuning sequences as 1\%, 5\%, 10\%, 20\% of the number of sequences in the original dataset. We have two observations from the results. First, the choice of the initial item has no significant influence on the watermark validity. Second, the watermark is valid when the number of fine-tuning sequences is small. The watermark validity drops significantly when the number of fine-tuning sequences reaches 20\%. However, this is an acceptable case, as it would be rarely possible for the attacker to obtain such a great number of data. Even if the attacker managed to do this, the model utility would drop significantly after fine-tuning. We show how the model utility is correlated with the watermark validity in \autoref{fig:validity vs utility}. Each point represents a model from \autoref{table:fine-tuning} where the initial item is the popular item. Other datasets produce similar results. For the model owner, the worst case is when the fine-tuned model has a high utility but a low watermark validity. This suggests that the attacker has successfully obtained a model with good performance and avoided the watermark detection. Such models would appear at the bottom right corner in \autoref{fig:validity vs utility}. However, we do not observe any point there. Instead, in \autoref{fig:validity vs utility}, the models with a high model utility also have a high watermark validity at the top right corner. On the other hand, those models that have a low watermark validity also have a low utility at the bottom left corner. Although we cannot detect the watermark in these models, they would be of no threat to the defender's target model since the utility is largely degraded. Therefore, AOW can effectively protect the target model from fine-tuning.

\begin{figure}[t]
\centering
    \includegraphics[width=0.45\textwidth]{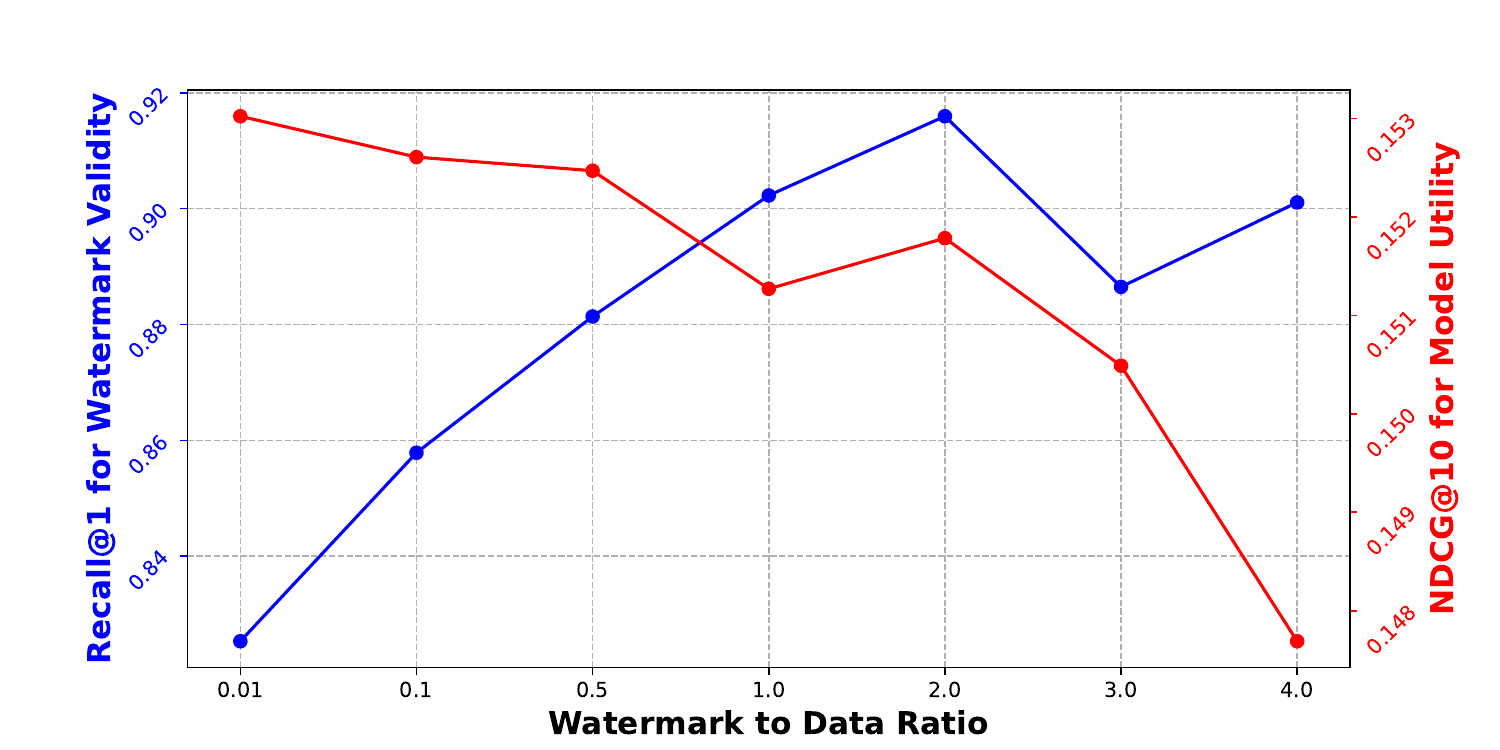}
\vspace{-1em}
\caption{Watermark-to-data ratio on Steam with the popular initial item and watermark length 20. The x-axis denotes the WDR ratio, which is the weight of the watermark sequence to the weight of the regular training set. The blue plot and the left y-axis denote the recall@1 of the watermark. The red plot and the right y-axis denote the NDCG@10 of the model utility.}
\label{fig:WDR}
\vspace{-1em}
\end{figure}

\subsection{RQ3: Hyperparameter Study}
In previous sections, we have reported results with different watermark length $n$ and compared the two different initial items, \textit{cold} and \textit{pop}. Therefore, we skip the study on the two hyperparameters and focus on the others in this section.
\subsubsection{Watermark-to-data Ratio}
\label{sec:wdr}
We show how the watermark-to-data ratio (WDR) influences the performance of AOW in \autoref{fig:WDR}. We show the results on Steam with the popular initial item and watermark length 20. We can observe that, as the WDR increases (the weight of $S_{wm}$ increases), the Recall@1 of the watermark also increases. However, the recall does not further increase significantly when WDR is larger than 1.0. Meanwhile, the NDCG@10 of the model decreases. Therefore, there is a trade-off between the watermark validity and the model utility as the WDR varies. However, the watermark validity is already at a very high level (recall@1>0.8) even if the WDR is low, so in order to preserve the utility of the model, we would suggest to select a low WDR. We by default set the WDR to 0.1 across all experiments.

\begin{figure}[t]
\centering
    \includegraphics[width=0.45\textwidth]{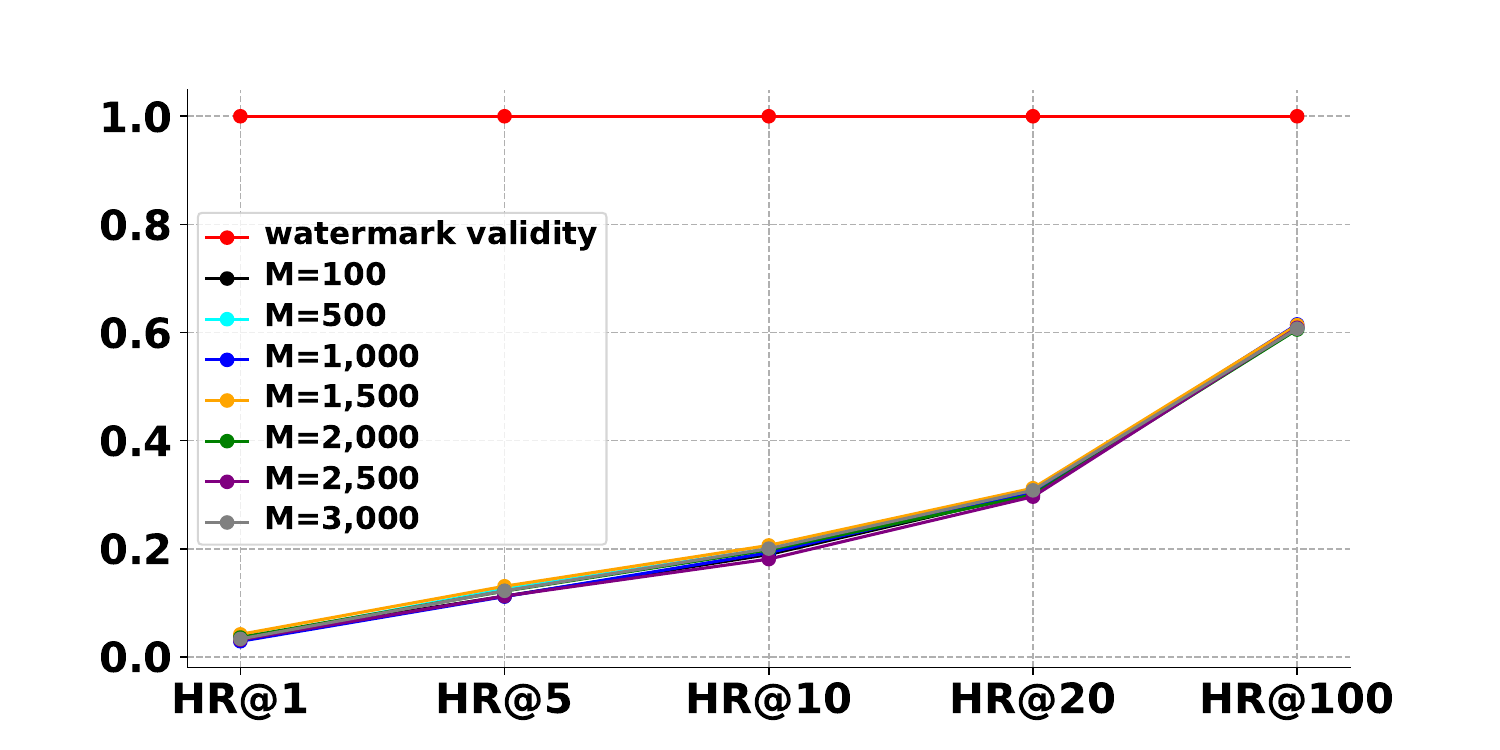}
\vspace{-1em}
\caption{Model utility and watermark validity with different choices of M in selecting the next item from the bottom-M items. The dataset is ML-1M. The watermark length is 20 and the initial item is the cold item. For different M, the watermark validity is consistently 100\%, which is represented by the red plot. Other plots represent the model utility of different M.}
\label{fig:M}
\vspace{-1em}
\end{figure}

\begin{table}[t]
\caption{Recall@100 in percentage for detecting the watermark in the oracle model of two different selection ranges for the next item in $S_{wm}$.}
\vspace{-1.5em}
\begin{center}
 \begin{tabular}{|c|c|c|} 
 \hline
  & Bottom-100 & Top-100 \\
 \hline
 Recall@100 & 0 & 34.02 \\
 \hline
\end{tabular}
\label{table:range}
\end{center}
\vspace{-2em}
\end{table}

\subsubsection{Bottom-M Item Selection Range}
\label{sec:M}
We show how the choice of M influences the model utility in \autoref{fig:M}. We use ML-1M and choose the cold item as the initial item, while setting the watermark length as 20. Other datasets and hyperparameter choices yield similar performance. It is clear that the choice of M does not introduce significant performance gap in terms of model utility. Besides, the watermark validity is consistently 100\% for all choices of M. 

However, it is not true if we claim that M does not matter. A larger M, in fact, may potentially harm the uniqueness of the watermark, causing it to be unintentionally detected in a benign oracle model. Consider two extreme cases, where we set the selection range of the next item in $S_{wm}$ as bottom-100 and top-100 respectively. We would like to see if we can detect the two watermarks from an oracle model. We set the initial item as the cold item, and the watermark length as 20. We train five oracle models with different random seeds on ML-1M, then generate 10 different watermark sequences from each oracle model and test them on other oracles. We report the average Recall@100 of the watermark in \autoref{table:range}. The Recall@100 of bottom-100 is zero, while it is 34.02\% for top-100. This suggests that incorporating top items into the selection range is risky as it would probably lead to a non-zero detection rate for the watermark in oracle models.

\subsubsection{Summary}
Here we provide a summary and recommended settings to all the hyperparameters. The watermark length is recommended to be as small as possible to ensure the memorization of it. For the initial item, cold items are generally a better choice since the model can memorize them easily because they are not as robust as the popular items. But popular items may become a better choice when the situations and demands change. WDR introduces a trade-off between the watermark validity and model utility. Setting it between 0.1 to 1.0 is generally a good choice to preserve model utility and ensure a high level of watermark validity. We can set the parameter M of bottom-M as 100 in most cases. But if a longer or diverse watermark is desired, M can be increased. In this case, we need to ensure that M has no overlap with the top-k evaluation range of the system to avoid the unintentional detection of the watermark in an oracle system.